\documentstyle[12pt]{article}
\begin{document}
\title
{Supersymmetry and Bogomol'nyi equations in the Abelian Higgs Model}
\author{
Jos\'e Edelstein
\thanks{CONICET}\\
Departamento de F\'\i sica, Universidad Nacional de La Plata
\thanks{Postal address: C.C. 67, (1900) La Plata, Argentina}\\
Argentina\\
\\
\rule{0cm}{0.8cm}
Carlos N\'u\~nez\\
Universidad de Buenos Aires\\
Argentina\\
\rule{0cm}{0.8cm}  and
\\
\\
Fidel Schaposnik
\thanks{Investigador CICBA, Argentina}\\
Departamento de F\'\i sica, Universidad Nacional de La Plata\\
Argentina}
\date{}
\maketitle

\def\thepage{\protect\raisebox{0ex}{\ } La Plata 93-07}
\thispagestyle{headings}
\markright{\thepage}

\begin{abstract}
The $N=2$ supersymmetric extension of the $2+1$ dimensional Abelian Higgs
model is discussed. By analysing the resulting supercharge algebra, the
connection between supersymmetry and Bogomol'nyi equations is clarified.
Analogous results are presented when the model is considered in
$2$-dimensional (Euclidean) space.
\end{abstract}

\newpage
\pagenumbering{arabic}

\section{Introduction}
Some twenty years ago Belavin, Polyakov, Schwartz and Tyupkin \cite{BPST}
found regular solutions to the four-dimensional Euclidean Yang-Mills
equations of motion (the well-honoured instantons) by showing that the
Yang-Mills action is bounded from below.
When the bound, which is of topological nature, is saturated, solutions
can be constructed by studying first order differential equations instead
of the more involved second order Euler-Lagrange equations.

Almost inmediately, other systems exhibiting first order differential
equations with solutions satisfying second order
equations of motion  were investigated \cite{dVS,Bogo}. They correspond
to gauge theories with spontaneous symmetry breaking and the
solutions were (static) solitons (vortices, monopoles)
saturating a bound of topological character, this time for the
energy.  More recently, Bogomol'nyi equations for the Chern-Simons-Higgs
system were also found \cite{JW}-\cite{Cor}.
Appart from the gauge coupling constant, symmetry breaking
implies the introduction of a second coupling constant for the Higgs
potential.
Strikingly enough, the topological bound and the resulting first-order
differential equations (known today as Bogomol'nyi equations) require certain
conditions on these coupling constants which reappear in other apparently
distinct fields.

Indeed, already in ref.\cite{dVS}, de Vega and one of the authors of the
present paper stressed that the required relation between coupling constants
in the Abelian Higgs model (which  corresponds
to the limit between type-I and type-II superconductivity in the Ginsburg
Landau model)
was precisely the same needed for its
supersymmetric extension \cite{Fayet}-\cite{SalSt}.

Although the connection between vortex solutions and supersymmetry in the
Abelian Higgs model
was afterwards thoroughly studied \cite{dVF}, the reasons
behind the overlap of these two
apparently divorced matters (supersymmetry and topological bounds) were not
investigated till very recently \cite{HS}-\cite{HSbis}. Much was
understood in these last works on the connection by analysing several models
but the case in
which the coincidence was first stressed, i.e. the Abelian Higgs model,
has not been detaily discussed yet.

It is the purpose of this work to study this issue, eventually finding
the reasons behind the coincident conditions imposed on the Abelian
Higgs model by supersymmetry and
the existence of Bogomol'nyi equations .
Originally, these first order equations were studied for static,
axially symmetric configurations. They could be interpreted as solitons
(Nielsen-Olesen vortices \cite{NO}).
In the approach of Hlousek-Spector   \cite{HS}-\cite{HSbis},
the identification of the
topological charge for the vortex configuration (its magnetic flux) with
the $N=2$ supersymmetry central charge plays a central r\^{o}le. Since
in the Abelian Higgs model
the topological charge is defined in 2-dimensional space, we shall first
work, taking advantage of axial symmetry, in
(2+1) space-time dimensions so that charges are given by
two-dimensional integrals.
Now, as it is well known, vortices in $(2+1)$ dimensions can be
interpreted as instantons in $2$-dimensional Euclidean space.
Although the general analysis of refs.\cite{HS}-\cite{HSbis} does not
apply to $2$-dimensional models, we shall  also study the
Abelian Higgs model as a bidimensional model showing that
again in this case the arguments connecting supersymmetry and Bogomol'nyi
equations can be applied.

Our results can be summarized as follows: starting from the Abelian
Higgs model in $2+1$ dimensions for which a topological charge can be found
and an $N=1$ supersymmetric extension can be constructed, an $N=2$
supersymmetric extension can be obtained provided a relation between
coupling constants holds.
The same relation is required for the existence of Bogomol'nyi equations.
Similar results hold when the Higgs model is defined in $2$ dimensional
Euclidean space.
The connection between $N=2$ supersymmetry and Bogomol'nyi equations is
clarified by the explicit construction of the supercharge algebra which
leads to a bound for the energy in terms of the central charge (which
coincides with the topological charge).
The bound is saturated by solutions of Bogomol'nyi equations.
In conclusion, in models with gauge symmetry breaking, $N=2$ supersymmetry
forces a relation between coupling constants and at the same time, through
its supercharge algebra, imposes Bogomol'nyi equations on field configurations.

\section{The Model}
The Abelian Higgs model dynamics in (2+1)-Minkowski space is defined by
the action:
\begin{equation}
{\cal S}_{\cal H} = \int d^3x \{-\frac{1}{4}F_{\mu\nu}F^{\mu\nu} +
\frac{1}{2}(D_{\mu}\phi)^*(D^{\mu}\phi) - \lambda(\vert\phi\vert^2 -
\phi_0^2)^2\}
\label{1}
\end{equation}
Here $\phi$ is a complex scalar,
\begin{equation}
\phi = \phi^1 + i\phi^2,
\label{2}
\end{equation}
the covariant derivative is defined as
\begin{equation}
D_{\mu}= \partial_{\mu} + ieA_{\mu}
\label{3}
\end{equation}
and $g^{\mu\nu}=(+ - -)$. An $N=1$ supersymmetric extension of this model
is given by the action:
\begin{eqnarray}
{\cal S}_{N=1} & = & \int d^3x \{ -\frac{1}{4}F_{\mu\nu}F^{\mu\nu} +
\frac{1}{2}(\partial_{\mu}M)(\partial^{\mu}M) + \frac{1}{2}
(D_{\mu}\phi)^*(D^{\mu}\phi) - 2\lambda M^2\vert\phi\vert^2 \nonumber \\
& - & \lambda (\vert\phi\vert^2 -
{\phi_0}^2)^2
+ \frac{i}{2}\overline{\rho}\not\!\partial\rho +
\frac{i}{2}\overline{\chi}\not\!\partial\chi +
\frac{i}{2}\overline{\psi}\not\!\! D\psi -
(2\lambda)^{1/2}M\overline{\psi}\psi \nonumber \\
& + & \frac{ie}{2}(\overline{\psi}\rho\phi - \overline{\rho}\psi\phi^*)
- (2\lambda)^{1/2}(\overline{\psi}\chi\phi + \overline{\chi}\psi\phi^*)\}
\label{4}
\end{eqnarray}
Our conventions for $\gamma$-matrices, (${\gamma^{\mu})_{\alpha}}^{\beta}$
are,
\begin{equation}
\gamma^0 = \left( \begin{array}{rr} 0 & -i \\ i & 0 \end{array} \right)
\; \; \; \; \;
\gamma^1 = \left( \begin{array}{rr} 0 & i \\ i & 0 \end{array} \right)
\; \; \; \; \;
\gamma^2 = \left( \begin{array}{rr} i & 0 \\ 0 & -i \end{array} \right)
\label{5}
\end{equation}
\[ \gamma^{\mu}\gamma^{\nu} = g^{\mu\nu} +
i\epsilon^{\mu\nu\lambda}\gamma_{\lambda}  \]
In (\ref{4}) $\rho$ and $\chi$ are Majorana fermions, $\psi$ is a Dirac
fermion and $M$ is a real scalar field. In fact, action (\ref{4}) can be
constructed by considering a complex scalar superfield $\Phi=(\phi,\psi,F)$,
a real spinor superfield $\Gamma^{\alpha}=(A_{\mu},\rho,\xi,\varphi)$ and
a real superfield $S=(M,\chi,D)$. Here $F$ and $D$ are auxiliary fields
which can be eliminated using their equations of motion. Concerning
$\xi$ and $\varphi$ they are absent in the Wess-Zumino gauge which we
adopt from here on.
It is easily seen that action (\ref{4}) is invariant under the
following supersymmetry transformations:
\begin{eqnarray}
\delta \rho & = & -\frac{i}{2}\epsilon^{\mu\nu\lambda}F_{\mu\nu}
\gamma_{\lambda}\eta \; \; \; , \; \; \; \delta A_{\mu} =
-i\overline{\eta}\gamma_{\mu}\lambda \; \; \; , \; \; \; \nonumber
\delta \psi = -i\gamma^{\mu} \eta D_{\mu}\phi - (8\lambda)^{1/2}M\phi\eta \\
\delta \chi & = & - (2\lambda)^{1/2}(\vert\phi\vert^2 - {\phi_0}^2)\eta -
i\gamma^{\mu}\eta\partial_{\mu}M  \; \; \; , \; \; \; \delta M =
\overline{\eta}\chi  \; \; \; , \; \; \; \delta \phi = \overline{\eta}\psi
\label{6}
\end{eqnarray}
where $\eta$ is an infinitesimal Majorana spinor.
Note that supersymmetry is achieved without imposing a relation between
$e$ and $\lambda$ as required in refs.\cite{Fayet,SalSt}.

The corresponding spinor supercurrent ${\cal J}_{N=1}^{\mu}$ is given by:
\begin{eqnarray}
{\cal J}_{N=1}^{\mu} & = & \frac{1}{4}\overline{\rho}\gamma^{\mu}
\epsilon^{\nu\lambda\sigma}F_{\nu\lambda}\gamma_{\sigma} -
i\left({\frac{\lambda}{2}}\right)^{1/2}\overline{\chi}\gamma^{\mu}
(\vert\phi\vert^2 - {\phi_0}^2) +
\frac{1}{2}\overline{\chi}\gamma^{\mu}\not\!\partial M \nonumber \\
& + & \frac{1}{2}\overline{\psi}\gamma^{\mu}\not\!\! D\phi
- i\overline{\psi}\gamma^{\mu}(2\lambda)^{1/2}M\phi
\label{7}
\end{eqnarray}

We now want to impose the requirement that the theory be invariant under
an $N=2$ extended supersymmetry.
This can be achieved by considering transformations (\ref{6}) with
complex parameter $\eta_c$ (now an infinitesimal Dirac spinor).
Being $\rho$ and $\chi$ real fermions,
we combine them into a Dirac
fermion $\Sigma$,
\begin{equation}
\Sigma \equiv \chi - i\rho
\label{8}
\end{equation}
so that, transformations (\ref{6}) become
\begin{eqnarray}
\hat{\delta} \Sigma & = &
- \left(\frac{1}{2}\epsilon^{\mu\nu\lambda}F_{\mu\nu}
\gamma_{\lambda} + (2\lambda)^{1/2}(\vert\phi\vert^2 - {\phi_0}^2) +
i\not\!\partial M\right) \eta_c
\; \; \; , \; \; \; \hat{\delta} A_{\mu} = -i\overline{\eta}_c\gamma_{\mu}
\lambda \nonumber  \\
\hat{\delta} \psi & = & -i\gamma^{\mu} D_{\mu}\phi \eta_c-
(8\lambda)^{1/2}M\phi \eta_c\; \; \; , \; \; \; \hat{\delta} M =
\overline{\eta}_c\chi  \; \; \; , \; \; \;
\hat{\delta} \phi = \overline{\eta}_c\psi
\label{9}
\end{eqnarray}

Using (\ref{8}), action (\ref{4}) can be rewritten in the form
\begin{eqnarray}
{\cal S}_{N=1} & = & \int d^3x \{ -\frac{1}{4}F_{\mu\nu}F^{\mu\nu} +
\frac{1}{2}(\partial_{\mu}M)(\partial^{\mu}M) + \frac{1}{2}
(D_{\mu}\phi)^*(D^{\mu}\phi) - 2\lambda M^2\vert\phi\vert^2 \nonumber \\
& - & \lambda (\vert\phi\vert^2 -
{\phi_0}^2)^2
+ \frac{i}{2}\overline{\Sigma}\not\!\partial\Sigma +
\frac{i}{2}\overline{\psi}\not\!\! D\psi -
(2\lambda)^{1/2}M\overline{\psi}\psi \nonumber \\
& - & \frac{e+(8\lambda)^{1/2}}{4}(\overline{\psi}\Sigma\phi + h.c.)
+ \frac{e-(8\lambda)^{1/2}}{4}(\overline{\psi}\Sigma^c\phi + h.c.)
\label{10}
\end{eqnarray}
Here ${\Sigma}^c$ is the charge conjugate (the complex conjugate)
of $\Sigma$.

Now, transformations (\ref{9}) with complex parameter $\eta_c = \eta
e^{-i\alpha}$ are equivalent to transformations (\ref{6}) with real parameter
$\eta$ followed by a phase transformation for fermions
$\Sigma$  ($\Sigma\to e^{i\alpha}\Sigma$) and $\psi$
($\psi\to e^{i\alpha}\psi$).
Then, $N=2$ supersymmetry requires invariance under this fermion
rotation.
One can easily see that fermion phase rotation invariance is achieved
if and only if:
\begin{equation}
\lambda = \frac{e^2}{8}
\label{11}
\end{equation}
Indeed, under a phase rotation, the factor $\overline{\psi}{\Sigma}^c\phi$
($[\overline{\psi}{\Sigma}^c\phi]^{\dag}$) appearing in the last
term in (\ref{10}) picks a $e^{-2i\alpha}$
($e^{+2i\alpha}$) phase, so that condition (\ref{11}) is needed in
order to make the last term to vanish. In this way phase rotation invariance
is ensured or, what is the same,
$N=2$ supersymmetry invariance is achieved.
This corresponds to a well-known result holding in general when,
starting from an $N=1$ supersymmetric gauge model, one attemps to impose
a second supersymmetry \cite{Sohn}:
conditions on coupling constants have to be imposed so as to accommodate
different $N=1$ multiplets into an $N=2$ multiplet.
Eq.(\ref{11}) is an example of such a condition.
In fact, condition (\ref{11})was obtained by
di Vecchia and Ferrara \cite{dVF} when looking for the $N=2$ supersymmetry
extension of the \underline{2-dimensional} Abelian Higgs model.

In summary we have arrived to the following $N=2$ supersymmetric action
associated with the Abelian Higgs model:
\begin{eqnarray}
{\cal S}_{N=2} & = & \int d^3x \{ -\frac{1}{4}F_{\mu\nu}F^{\mu\nu} +
\frac{1}{2}(\partial_{\mu}M)(\partial^{\mu}M) + \frac{1}{2}
(D_{\mu}\phi)^*(D^{\mu}\phi) - \frac{e^2}{4}M^2\vert\phi\vert^2 \nonumber \\
& - & \frac{e^2}{8}(\vert\phi\vert^2 -
{\phi_0}^2)^2
+ \frac{i}{2}\overline{\Sigma}\not\!\partial\Sigma +
\frac{i}{2}\overline{\psi}\not\!\! D\psi -
\frac{e}{2} M\overline{\psi}\psi \nonumber \\
& - & \frac{e}{2}(\overline{\psi}\Sigma\phi + h.c.)
\label{10bis}
\end{eqnarray}
In the next section, the reasons why condition (\ref{11}) ensuring $N=2$
supersymmetry is also needed in order to attain the Bogomol'nyi bound
will be clear at the light of Hlousek-Spector approach \cite{HS}-\cite{HSbis}.

\section{Supercharge and Bogomol'nyi equations}
$N=2$ supercharges generating the supersymmetry transformations (\ref{9})
can be found from the conserved quantity
\begin{equation}
{\cal Q} = \frac{\sqrt 2}{e\phi_0}\int d^2x {\cal J}_{N=2}^0
\label{12}
\end{equation}
where ${\cal J}_{N=2}^{\mu}$ is the Noether current associated with
transformations (\ref{9}). Writing
\begin{equation}
{\cal Q} = \overline{\eta}_cQ + \overline{Q}\eta_c
\label{13}
\end{equation}
we find
\begin{eqnarray}
Q & = & \frac{\sqrt 2}{e\phi_0}\int d^2x [\left(
-\frac{1}{2}\epsilon^{\mu\nu\lambda}
F_{\mu\nu}\gamma_{\lambda} + i\not\!\partial M - \frac{e}{2}
(\vert\phi\vert^2 - {\phi_0}^2) \right) \gamma^0\Sigma \nonumber \\
& + & \left( i(\not\!\! D\phi)^* - \frac{e}{2}M\phi^* \right)
\gamma^0 \psi ]
\label{14}
\end{eqnarray}
and
\newpage
\begin{eqnarray}
\overline{Q} & = & \frac{\sqrt 2}{e\phi_0}\int d^2x [
\overline{\Sigma}\gamma^0\left(-
\frac{1}{2}\epsilon^{\mu\nu\lambda}F_{\mu\nu}\gamma_{\lambda} -
i\not\!\partial M - \frac{e}{2}(\vert\phi\vert^2 - {\phi_0}^2)
\right) \nonumber \\
& + & \overline{\psi}\gamma^0 \left( -i\not\!\! D\phi -
\frac{e}{2}M\phi \right) ]
\label{15}
\end{eqnarray}

We are interested in connecting the supercharge algebra and
Bogomol'nyi equations for vortices of the original Abelian Higgs model.
We then restrict the model to its bosonic sector by putting $M = 0$ and,
after computing the supercharge algebra,
all fermions to zero so as to end with the
original Abelian Higgs model.
Moreover, since Bogomol'nyi equations correspond to static configurations
with $A_0=0$, we impose these conditions on
(\ref{14})-(\ref{15}), finding for the anticommutation relation among
spinor supercharges $Q$ , $\overline{Q}$:
\begin{equation}
\{Q_{\alpha},\overline{Q}^{\beta}\} = 2{(\gamma_{0})_{\alpha}}^{\beta}P^{0}
+ {\delta_{\alpha}}^{\beta} T
\label{16}
\end{equation}
where
\begin{equation}
P^0 = E = \frac{1}{2 e^2{\phi_0}^2}
\int d^2x \left[ \frac{1}{2}F^2_{ij} + \vert D_i\phi\vert^2
+ \frac{e^2}{4}(\vert\phi\vert^2 - {\phi_0}^2)^2 \right]
\label{17}
\end{equation}
while the central charge is given by:
\begin{equation}
T = -\frac{1}{e^2{\phi_0}^2}\int d^2x
\left[ \frac{e}{2}\epsilon^{ij}F_{ij}(\vert\phi\vert^2 - {\phi_0}^2)
+ i\epsilon^{ij}(D_i\phi)(D_j\phi)^* \right]
\label{18}
\end{equation}
Here $i,j=1,2$.

We are now ready to find the connection between $N=2$ supersymmetry and
Bogomol'nyi equations. Let us start by noting that the central charge in
(\ref{18}) coincides with the topological charge of the Abelian Higgs model.
Indeed, $T$ can be rewritten in the form
\begin{equation}
T = \int \partial_i{\cal V}^i d^2x
\label{18bis}
\end{equation}
where ${\cal V}^i$ is given by
\begin{equation}
{\cal V}^i =  \left(\frac{1}{e}A_j + \frac{i}{e^2{\phi_0}^2}
{\phi}^*D_j\phi \right)\epsilon^{ij}
\label{18bisbis}
\end{equation}
so that, after Stokes theorem
(and taking into account that $D_i\phi \to 0$ at infinity)
\begin{equation}
T = \frac{1}{e}  \oint A_idx^i = \frac{2\pi n}{e}
\label{18bisbisbis}
\end{equation}
where $n, n \in Z$, is the integer characterizing the homotopy class to
which $A_{i}$ belongs.

This sort of identity between the $N=2$ central charge and topological charge
was first stressed by Olive and Witten \cite{OW} in their study of the
$SO(3)$ Georgi-Glashow model in the Prasad-Sommerfield limit. It was also
discussed for the self-dual Chern-Simons system by Lee, Lee and Weinberg
\cite{LLW}. More recently, Hlousek and Spector \cite{HS}-\cite{HSbis}
have thoroughly
analysed this conection by studying several models where the existence
of an $N=1$ supersymmetry and a topological current implies an $N=2$
supersymmetry with its central charge coinciding with the topological
charge. It is important to stress at this point that in order to
achieve the $N=2$ supersymmetry one is forced to impose the
condition (\ref{11}) exactly as it happens when trying to find a Bogomol'nyi
bound for the Abelian Higgs model. This condition is unavoidable both for
having $N=2$ supersymmmetry and Bogomol'nyi equations.
Also, in the study of self-dual Chern-Simons systems, for which a topological
charge (related to the magnetic flux) and an $N=1$ extension does exist,
a condition on the symmetry breaking coupling
constant must be imposed both to achieve $N=2$ extended
supersymmetry \cite{LLW} and to
obtain Bogomolnyi equations \cite{JW},\cite{Cor}.

At the light of the discussion above,  it should be worthwhile  to
formulate the result in \cite{HS}-\cite{HSbis}, when applied to models with
symmetry breaking by stating that:

{}~

\noindent { For gauge theories with spontaneous symmetry breaking and
a topological charge with an $N=1$ supersymmetric version,
the N=2 supersymmetric extension, which requires certain conditions
for coupling constants, has a central charge coinciding with the
topological charge}

{}~

It is now easy to find \cite{HS}-\cite{HSbis} the Bogomol'nyi bound from the
supersymmetry algebra (\ref{16}).
Indeed, since the anticommutators in
(\ref{16}) are Hermitian, one has:
\begin{equation}
\{Q_{\alpha},\overline{Q}^{\beta}\}\{Q^{\alpha},\overline{Q}_{\beta}\}
\geq 0
\label{19}
\end{equation}
or using (\ref{16}),
\begin{equation}
E \geq \vert T \vert
\label{20}
\end{equation}
Now, in view of (\ref{17}) and (\ref{18}), one finds Bogomol'nyi result:
\begin{equation}
E \geq \frac{2\pi}{e}\vert n \vert
\label{21}
\end{equation}

In order to explicitly obtain Bogomol'nyi equations (saturating the
energy bound) from the supersymmetry
algebra, we define, following Hlousek and Spector \cite{HS}-\cite{HSbis}:
\begin{equation}
Q_{I} = \frac{Q_+ + i Q_-}{\sqrt{2}}
\label{22}
\end{equation}
\begin{equation}
Q_{II} = \frac{\overline{Q}^+ + i\overline{Q}^-}{\sqrt{2}}
\label{23}
\end{equation}
where we have defined $Q_{\pm}$ from
\begin{equation}
Q = \left( \begin{array}{c} Q_+ \\ Q_- \end{array} \right)
\label{24}
\end{equation}
\begin{equation}
\overline{Q} = \left( \overline{Q}^+ \; \; \; \overline{Q}^- \right)
\label{25}
\end{equation}
Now, suppose that a field configuration $\vert B>$  saturates the
Bogomol'nyi bound derived from (\ref{19}). Then, one necessarily has
\begin{equation}
\left( Q_{I} \pm Q_{II} \right) \vert B> = 0
\label{26}
\end{equation}
or, using (\ref{22})-(\ref{25}) and (\ref{14})-(\ref{15})
\begin{eqnarray}
\epsilon^{ij}F_{ij} & = & \pm {e} (\vert\phi\vert^2 -
{\phi_0}^2) \nonumber \\
i\epsilon_{ij}D^i\phi & = & \pm (D_j\phi)^*
\label{27}
\end{eqnarray}
These are of course the Bogomol'nyi equations for the Abelian Higgs
model \cite{dVS}-\cite{Bogo}.
Due to (\ref{21}), their solution also solves the static
Euler-Lagrange equations of motion. Let us insist that the condition
(\ref{11}) necesary for this last fact, arises in the present approach
from the requirement of $N=2$ supersymmetry.

In conclusion, in models with gauge symmetry breaking,
like the Abelian Higgs model, $N=2$ supersymmetry
forces a relation between coupling constants and at the same time, through
its supercharge algebra, imposes Bogomol'nyi equations on field configurations.

\section{The Model in d=2 dimensions}
The Hlousek-Spector \cite{HS}-\cite{HSbis}
approach was basically formulated in ($2+1$) and
($3+1$) dimensions so that our ($2+1$) discussion above enters in this
family of models.
Concerning ($1+1$) models, although no general results are
presented in refs.\cite{HS,HSbis}, connection between supersymmetry and
Bogomol'nyi equations are well-known
in this case \cite{dVF}-\cite{OW},\cite{HS}.
We would like to show in this section that results analogous to those
obtained in section 3 can be found if one considers the Abelian Higgs
model in ($1+1$) Euclidean dimensions where vortex configurations
become instantons.

Following di Vecchia and Ferrara \cite{dVF}, the $N=1$ supersymmetry model
to consider in $d=2$ dimensions is (in the Wess-Zumino gauge)
\begin{eqnarray}
S_{N=1} & = & \int d^2x \{ - \frac{1}{4}F_{\mu \nu}^2
+ \frac{1}{2}(\partial_{\mu} P)^2 - \frac{1}{2}(\partial_{\mu} M)^2
- \frac{1}{2}({\cal D}_{\mu} \vec{\Phi})^2
+ \frac{1}{2}e^2 P^2 \vec{\Phi}^2
\nonumber \\
& + & \frac{i}{2} \rho\not\!\partial\rho
+ \frac{i}{2}\vec{\psi}\cdot\not\!\!{\cal D}\vec{\psi}
+ \frac{i}{2}\chi\not\!\partial\chi
- \frac{ie}{2}\vec\psi\wedge\vec\psi P +
ie\vec\psi\wedge\gamma_5\rho\vec{\Phi} \\
& - & i\sqrt{2\lambda}M\vec\psi\cdot\gamma_5\vec\psi
- i2\sqrt{2\lambda}\vec{\Phi}\cdot\vec\psi\gamma_5\chi -
4\lambda M^2\vec{\Phi}^2 - \lambda
(\vec{\Phi}^2 - {\Phi_0}^2)^2 \} \nonumber
\label{28}
\end{eqnarray}
where
$\vec{\Phi}$ is the (real) Higgs field doublet
which is in the adjoint representation
of $SO(2)$, $\vec{\psi}$ is a doublet of Majorana fermions,
$\rho$ and $\chi$ are Majorana fermions and
$M$ ($P$) is a real scalar (pseudoscalar).
The $\gamma$-matrices are taken as:
\[ \gamma_0 = \left( \begin{array}{rr} 1 & 0 \\ 0 & -1 \end{array} \right)
\; \; \; \; \;
\gamma_1 = \left( \begin{array}{rr} 0 & 1 \\ 1 & 0 \end{array} \right)
\; \; \; \; \;
\gamma_5 = \left( \begin{array}{rr} 0 & 1 \\ -1 & 0 \end{array} \right) \]
It is understood that $\vec{\psi}\cdot\vec{\Phi}$ means $\psi^a\Phi^a$
while $\vec{\psi}\wedge\vec{\Phi}$ is the (pseudo)scalar
$\epsilon^{ab}\psi^a\Phi^b$.

It is interesting to note that action (\ref{28}) can be obtained by
dimensional reduction (including identification of $A_0$ in $d=3$
with $P$ in $d=2$) of action (\ref{4}).
Hence, $N=1$ supersymmetry transformations can simply be inferred
from transformations (\ref{6}):
\begin{eqnarray}
\delta \Phi^a & = & i\epsilon\psi^a
 \; \; \; , \; \; \;
\delta \chi = (\not\!\partial M - \sqrt{2\lambda}(\vec{\Phi}^2
- {\Phi_0}^2)\gamma_5 )\epsilon \nonumber \; \; \; ,
\; \; \;
\delta P = - \frac{i}{2}\epsilon\gamma_5\rho  \\
\delta A_{\mu} & = & - \frac{i}{2}\epsilon\gamma_5\gamma_{\mu}\rho
\; \; \; , \; \; \;
\delta \psi^a = (\not\!\!{\cal D}\Phi^a
- 2\sqrt{2\lambda}\gamma_5 M \Phi^a
+ eP\epsilon^{ab} \Phi^b)\epsilon \nonumber \\
\delta \rho & = & (\gamma_5\not\!\partial P
- \frac{1}{2}\epsilon_{\mu \nu}F_{\mu \nu})\epsilon
\; \; \; , \; \; \; \delta M = i\epsilon\chi
\label{29}
\end{eqnarray}

It is evident that condition (\ref{11}) must hold also in the present
$d=2$ model in order to have $N=2$
extended supersymmetry, as was first
pointed out by di Vecchia-Ferrara \cite{dVF}.
As discussed previously, it ensures that fermions $\chi$ and $\rho$
can be accommodated into a unique $N=2$ supermultiplet.
If we call $\Sigma^1 = \rho$ and $\Sigma^2 = \gamma_5\chi$,
and impose condition
(\ref{11}), the resulting $N=2$ supersymmetry transformations
for any field $\phi$ read
\begin{equation}
\hat{\delta} \phi = \delta_1\phi \eta^1 +
\delta_2\phi\eta^2
\label{29bis}
\end{equation}
with $\eta_c = \eta_1+i \eta_2$, the $N=2$ transformation parameter.
In component fields, eq.(\ref{29bis})
is:
\begin{eqnarray}
\delta_1 M & = & -i\gamma_5\Sigma^2 \; \; \; , \;\;\;
\delta_2 M = -i\gamma_5\Sigma^1 \; \; \; , \;\;\;
\delta_1 \Phi^a = i\psi^a    \;\;\; , \; \;\;
\delta_2 \Phi^a = -i\epsilon^{ab}\gamma_5\psi^b   \nonumber \\
\delta_1 A_{\mu} & = & -i\gamma_5\gamma_{\mu}\Sigma^1 \; \; \; , \;\;\;
\delta_2 A_{\mu} = -i\gamma_5\gamma_{\mu}\Sigma^2    \;\;\; , \;\;\;
\delta_a P = -i\gamma_5\Sigma^a \nonumber \\
\delta_1\psi^a & = & \not\!\!{\cal D}\Phi^a + eP\gamma_5\Phi^a -
eM\epsilon^{ab}\Phi^b        \nonumber                    \\
\delta_2\psi^a & = & \epsilon^{ab}\gamma_5\not\!\!{\cal D}\Phi^b
- eM\gamma_5\Phi^a + eP\epsilon^{ab}\Phi^b    \nonumber            \\
\delta_a\Sigma^a & = & -(-1)^a\gamma_5\not\!\partial P -
\frac{1}{2}\epsilon_{\mu\nu}F_{\mu\nu} \nonumber \\
\delta_a\Sigma^b & = & \epsilon^{ab}\gamma_5\not\!\partial M
+ \frac{e}{2}(\vec{\Phi}^2 - {\Phi_0}^2)
\label{30}
\end{eqnarray}
At this point let us note that conditions (\ref{26}) leading to
Bogomol'nyi equations, can be seen to be equivalent to:
\begin{eqnarray}
(\delta_1 \pm \delta_2)\vec{\psi} = 0 \nonumber \\
(\delta_1 \pm \delta_2)\vec{\Sigma} = 0
\label{32}
\end{eqnarray}
Now, it is easy to see that (\ref{32}) implies:
\begin{eqnarray}
P & = & \mp M \label{33}\\
\epsilon_{\mu\nu}F_{\mu\nu} & = & \pm
{e} (\vec{\Phi}^2 - {\Phi_0}^2) \label{34} \\
\epsilon_{\mu\nu}\epsilon^{ab}{\cal D}_{\nu}\Phi^b & = & \pm
{\cal D}_{\mu}\Phi^a  \label{35}
\end{eqnarray}

We can recognize eqs.(\ref{34})-(\ref{35}) as the usual self-dual
equations whose solutions saturate a bound of the action.
It can be seen that the other fields entering in the $ N=2$  supersymmetric
model also satisfy
Bogomol'nyi equation. Indeed,
following the same procedure, one can impose to bosonic fields:
\begin{equation}
(\delta_1 \pm \delta_2)\phi = 0
\label{36}
\end{equation}
so that one obtains:
\begin{eqnarray}
\rho & = & \pm \gamma_5\chi \label{37} \\
\psi^a & = & \pm \epsilon^{ab}\gamma_5\psi^b \label{38}
\end{eqnarray}
which are the fermionic superpartner of Bogomol'nyi equations appearing
above \cite{dVF}.

To conclude, we would like to point that the connection between
supersymmetry and self-duality equations
might be of interest in the study of gravity.
In particular, a $d=2$ model for gravity coupled to matter can be constructed
from a topological model where self-duality equations play a central role
\cite{Let1}-\cite{Let2}. In this context, as well as in the study of black
hole solutions \cite{Kal},
supersymmetry should provide a natural framework to analyse classical
and quantum properties.
We hope to report on these issues in a forthcoming work.

\end{document}